\documentclass[
 aps,pre,reprint,
 amsmath,amssymb,superscriptaddress
] {revtex4-1}

\usepackage{graphicx}
\usepackage{hyperref,amsmath}
\usepackage{braket,amsmath}
\usepackage{dcolumn}
\usepackage{bm}
\usepackage{newfloat}
\usepackage{multibib}
\usepackage{color}

\begin{document}

\title{Robust electric dipole transition at microwave frequencies for nuclear spin qubits in silicon}

\author{Guilherme Tosi}
 \affiliation{Centre for Quantum Computation and Communication Technology, School of Electrical Engineering \& Telecommunications, UNSW Sydney, New South Wales 2052, Australia.}
\author{Fahd A. Mohiyaddin}
 \altaffiliation[Present address: ]{Quantum Computing Institute, Oak Ridge National Laboratory, Oak Ridge, TN, USA}
 \affiliation{Centre for Quantum Computation and Communication Technology, School of Electrical Engineering \& Telecommunications, UNSW Sydney, New South Wales 2052, Australia.}
\author{Stefanie Tenberg}
 \affiliation{Centre for Quantum Computation and Communication Technology, School of Electrical Engineering \& Telecommunications, UNSW Sydney, New South Wales 2052, Australia.}
\author{Arne Laucht}
 \affiliation{Centre for Quantum Computation and Communication Technology, School of Electrical Engineering \& Telecommunications, UNSW Sydney, New South Wales 2052, Australia.}
\author{Andrea Morello}
 \email{a.morello@unsw.edu.au}
 \affiliation{Centre for Quantum Computation and Communication Technology, School of Electrical Engineering \& Telecommunications, UNSW Sydney, New South Wales 2052, Australia.}

\date{\today}

\begin{abstract}

The nuclear spin state of a phosphorus donor ($^{31}$P) in isotopically enriched silicon-28 is an excellent host to store quantum information in the solid state. The spin's insensitivity to electric fields yields a solid-state qubit with record coherence times, but also renders coupling to other quantum systems very challenging. Here, we describe how to generate a strong electric dipole ($>100$~Debye) at microwave frequencies for the nuclear spin. This is achieved by applying a magnetic drive to the spin of the donor-bound electron, while simultaneously controlling its charge state with electric fields. Under certain conditions, the microwave magnetic drive also renders the nuclear spin resonance frequency and electric dipole strongly insensitive to electrical noise, yielding long ($>1$~ms) dephasing times and robust gate operations. The nuclear spin could then be strongly coupled to microwave resonators, with a vacuum Rabi splitting of order 1 MHz, or to other nuclear spins, nearly half a micrometer apart, via strong electric dipole-dipole interaction. This work brings the $^{31}$P nuclear qubit into the realm of hybrid quantum systems and opens up new avenues in quantum information processing.

\end{abstract}

\maketitle

\section{Introduction}

The nuclear spin of a phosphorus donor in silicon has long been the subject of much study in the context of solid-state quantum information processing, either as a qubit cell for large-scale quantum processors \cite{Kane1998,Ogorman2016,Hill2015}, or a memory for long-lived quantum information storage \cite{Morton2008,Freer2017}. Whether in ensemble form \cite{Saeedi2013} or as individual qubit \cite{Muhonen2014}, the $^{31}$P nuclear spin has record-long coherence times, thanks to its insensitivity to electric fields and the possibility to drastically reduce magnetic environmental noise by hosting it in isotopically pure $^{28}$Si \cite{Itoh2014}. However, it cannot trivially be coupled to other quantum systems, and therefore all quantum computing proposals so far impose short interaction distances and slow quantum gate operations \cite{Kane1998,Ogorman2016,Hill2015}.

In the hybrid approach to quantum information processing \cite{Xiang2013}, different quantum systems interact in a large architecture that benefits from the best properties of each system, which are often coupled together via microwave resonators. In order to couple to individual spin qubits, the resonator vacuum field can be enhanced by shrinking its dimensions in the vicinity of the spin qubit, thereby enhancing the spin-photon coupling rate \cite{Tosi2014,Jenkins2016,Haikka2017,Sarabi2017}. However, having a Zeeman splitting in the radio-frequency range and a null electric dipole, phosphorus nuclear-spins do not interact naturally with microwave resonators.

The artificial creation of electric dipole transitions has been proposed for different spin systems \cite{Pioro-Ladriere2008,Shi2012,Russ2017,Salfi2016,Tosi2017} as a way to facilitate scalability. The challenge here is how to make the spin drivable by electric fields without making it too susceptible to electrical noise, which can be significant in nanoscale electronic devices. Here we show how to engineer a strong electric dipole transition at microwave frequencies for the nuclear spin, by applying an oscillating magnetic field and by sharing an electron between the donor and a quantum dot defined at the Si/SiO$_2$ interface \cite{Calderon2006,Veldhorst2014, Tosi2017,Harvey-Collard2017}. While the admixture of spin and charge states can potentially make the system very sensitive to electric noise, we show that the nuclear spin precession frequency and electric dipole strength can be rendered highly immune to electrical noise by a peculiar choice of spin-charge hybridization. By providing a robust coupling between the nuclear spin and electric fields, our scheme opens up new avenues to couple $^{31}$P qubits to other quantum systems, including microwave resonators, superconducting qubits, or simply other nuclear spins but at distances and with speeds that had not been anticipated so far.

The paper is organized as follows: in Sec.~\ref{sec:Raman} we describe the basic concept of a Raman-like nuclear spin control, utilizing a magnetic and an electric drive on the $^{31}$P donor system; in Sec.~\ref{sec:strong} we show how to make this Raman drive significantly stronger, by involving the electron charge state and creating a strong electric dipole transition, at the cost of tarnishing the nuclear spin coherence; in Sec.~\ref{sec:robust} we explain how the complete hybridization of all degrees of freedom -- nuclear spin, electron spin, and charge -- leads (rather non-trivially) to an even stronger electric dipole transition, while minimizing the sensitivity of the nuclear state to charge noise. Sec.~\ref{sec:long} contains a brief discussion of how to use these ideas for long-distance coupling of nuclear spin qubits, followed by a conclusion and outlook in Sec.~\ref{sec:conclusion}.

\begin{figure}
\centering
\includegraphics[width=0.9\columnwidth]{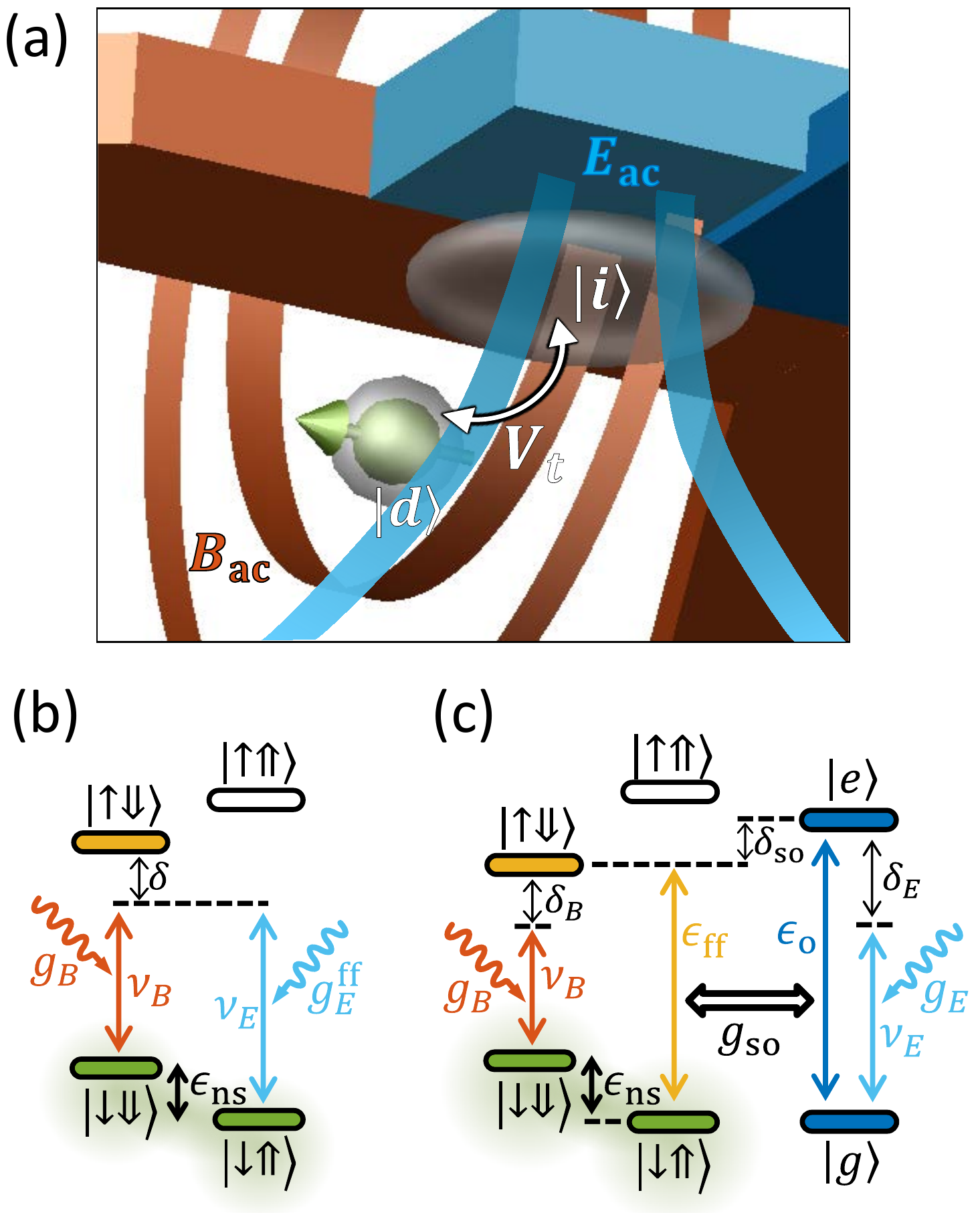}
\caption{
		(a) Components of a Raman-enabled Si:P nuclear electric dipole transition. The electron  spatial wavefunction (transparent gray) is shared between an interface-dot, $|i\rangle$, and a donor-bound state, $|d\rangle$, which are coupled by a tunnel rate $V_t$. Metallic gates (blue) on top of SiO$_2$ dielectric (not shown) control the electron charge state via a static vertical field $E_{\rm dc}$, and can introduce an oscillating electric field $E_{\rm ac}$. In a circuit-Quantum Electrodynamics setup, the electrostatic gate can be replaced by the inner conductor of a microwave resonator, and $E_{\rm ac}$ by the vacuum field of such resonator. A nearby broadband antenna \cite{Dehollain2013} (brown) provides the magnetic drive $B_{\rm ac}$. (b) Simplified energy level diagram for Raman-drive of the Si:P nuclear-spin qubit, with energy splitting $\epsilon_{\rm ns}$. $\ket{\uparrow}$ ($\ket{\downarrow}$) represents electron spin up (down), while $\ket{\Uparrow}$ ($\ket{\Downarrow}$) represents nuclear spin up (down). The second-order Raman drive is obtained by combining the microwave electric and magnetic drives, having frequencies $\nu_B$ and $\nu_E$, respectively, and coupling rates $g_B$ and $g_E^{\rm ff}$, respectively. The drive is detuned by a frequency $\delta$ from the $\ket{\uparrow \Downarrow}$ state. (c) Expanded energy diagram including the charge states $\ket{g}$ and $\ket{e}$.
}
\label{fig:Raman}
\end{figure}

\section{Second-order Raman drive of a $^{31}$P nuclear spin} \label{sec:Raman}

The key ingredients of our proposal are illustrated in Fig.~\ref{fig:Raman}a. The $^{\rm 31}$P nuclear spin ($I=1/2$) interacts with the spin ($S=1/2$) of a donor-bound electron via the contact hyperfine interaction, described by the Hamiltonian (in frequency units, i.e. setting the Planck constant $h=1$):

\begin{equation}
\mathcal{H}_A=A\ {\bf S\cdot I},
\end{equation}

where $\textbf{S}$ and $\textbf{I}$ are the electron and nucleus spin operators, respectively, and $A$ is the electron-nuclear hyperfine coupling ($A\approx117$~MHz in bulk silicon) arising from Fermi contact interaction. Further, we assume that a static magnetic field $B_0$ is applied to the donor and splits the electron ($\ket{\uparrow}, \ket{\downarrow}$) and nuclear ($\ket{\Uparrow}, \ket{\Downarrow}$) spin states via the Zeeman Hamiltonian:

\begin{equation}
\mathcal{H}_{B_0}=B_0\left(\gamma_eS_z-\gamma_nI_z\right),
\end{equation}

where $\gamma_e\approx28$~GHz/T and $\gamma_n\approx17.2$~MHz/T are the electron and nuclear gyromagnetic ratios, respectively. Fig. \ref{fig:Raman}b shows the donor energy levels (not to scale) when $\gamma_eB_0\gg A$, in which case the eigenstates of the Hamiltonian are well approximated by the tensor product of the electron and nuclear basis states.

The electron and nuclear spins can be coherently driven by conventional magnetic resonance transitions using oscillating magnetic fields at microwave \cite{Pla2012} and radio frequencies \cite{Pla2013}. In particular, the nuclear spin transition frequency when the electron spin is $\ket{\downarrow}$ (i.e. the $\ket{\downarrow \Uparrow} \leftrightarrow \ket{\downarrow \Downarrow}$ transition) is:
\begin{equation} \label{eq:epsilon_ns}
\epsilon_{\rm ns}(A) = \gamma_nB_0 + A/2.
\end{equation}

However, we have also recently proposed \cite{Tosi2017} a new electrically-driven spin transition between the states $\ket{\downarrow\Uparrow}\leftrightarrow\ket{\uparrow\Downarrow}$. The qubit that results from adopting  $\ket{\downarrow\Uparrow}, \ket{\uparrow\Downarrow}$ as basis states is called the `flip-flop' qubit, with energy splitting

\begin{equation}
\epsilon_{\rm ff} = \sqrt{(\gamma_+ B_0)^2 + A^2},
\end{equation}

where $\gamma_+ = \gamma_e + \gamma_n$. The hyperfine coupling $A$ can be modified from its bulk value by the application of a vertical electric field $E_z$, which displaces the electron from the nucleus \cite{Kane1998,Laucht2015,Tosi2017}. If $E_z$ also contains an oscillating component $E_{\rm ac}$ at microwave frequency $\nu_E\approx \gamma_+ B_0$, the hyperfine coupling acquires a time-dependent component $A_{\rm ac}\cos{\left(2\pi\nu_E t\right)}$ which renders $\mathcal{H}_A(t)$ time-dependent as well. Since $\mathcal{H}_A(t)$ is a transverse term in the flip-flop subspace, this results in electrically driving transitions between the flip-flop states at a rate\cite{coupling}:

\begin{equation}
g_E^{\rm ff}=A_{\rm ac}/4.
\end{equation}

Now we show how the flip-flop transition provides a way of controlling the nuclear spin state without any radiofrequency field, by using instead two microwave-frequency excitations, one of which is electric. This has important advantages over magnetic-only schemes \cite{Morton2008,Freer2017}, since it allows coupling a nucleus to the vacuum electric field of a microwave cavity, or to another nucleus similarly equipped with an electric dipole, as we will show below. The key idea is to combine the electrical drive of the flip-flop transition with an additional magnetic drive $B_{\rm ac}\cos(2\pi\nu_Bt)$, perpendicular to the static $B_0$ (Fig \ref{fig:Raman}a). This is the conventional electron spin resonance (ESR) transition, that couples the electron spin states $\lvert\downarrow\Downarrow\rangle$ and $\lvert\uparrow\Downarrow\rangle$ at a rate:

\begin{equation}
g_{B}=\gamma_eB_{\rm ac}/4.
\end{equation}

Combining these two driving fields results in a process analogous to a Raman transition \cite{Kok2010}, as shown in Fig. \ref{fig:Raman}b: with the electron in the ground spin state $\ket{\downarrow}$, the AC electric and magnetic fields drive the nuclear-spin ``up'', $\lvert\downarrow\Uparrow\rangle$, and ``down'', $\lvert\downarrow\Downarrow\rangle$, states, respectively, to a virtual level detuned from the $\lvert\uparrow\Downarrow\rangle$ state by $\delta\gg g_B,g_E^{\rm ff}$. As a result, the nuclear spin is driven via a second order process, with minimal excitation of the electron spin, at a rate:

\begin{equation} \label{eq:g_E^ns_simple}
g_E^{\rm ns}=\frac{g_B g_E^{\rm ff}}{\delta}.
\end{equation}

This can be interpreted as meaning that the microwave magnetic drive $B_{\rm ac}$ creates an electric dipole transition for the nuclear spin, with strength:

\begin{equation}
p_E^{\rm ns} = \frac{4 g_E^{\rm ns}}{E_{\rm ac}} = g_B \frac{A_{\rm ac}}{E_{\rm ac} \delta}.
\label{eq:pns}
\end{equation}

The ESR transition rate $g_B$ can reach a few MHz \cite{Pla2012} with realistic values of the oscillating magnetic field, but the overall rate $g_E^{\rm ns}$ and the associated electric dipole $p_E^{\rm ns}$ depend also on $A_{\rm ac}$, i.e. on the degree in which the oscillating electric field $E_{\rm ac}$ is capable of affecting the donor electron wavefunction. If the electron is mostly confined at the donor Coulomb potential, $A$ is quite insensitive to electric fields. Estimating $A_{\rm ac}/E_{\rm ac}$ from the value $\partial A/\partial E\approx0.5~{\rm Hz\cdot m/V}$ extracted in a recent experiment \cite{Laucht2015}, the corresponding Raman-enabled nuclear electric dipole would be only $p_E^{\rm ns} \approx 10~\mu$Debye (assuming $\delta=10g_B$) \footnote{Since 1 Debye $= 3.336 \times 10^{-30}$ C$\cdot$m, the electric dipole moment expressed in units of Hz$\cdot$m/V is converted to Debye via the factor $h/3.336 \times 10^{-30} = 1.986 \times 10^{-4} \approx 1/5000$.}.

\section{Strong electric dipole transition of a $^{31}$P nuclear spin} \label{sec:strong}

To increase the nuclear spin electric dipole transition strength, the $\partial A/\partial E$ needs to be strongly enhanced. This can be achieved by placing the donor underneath a quantum dot located at the Si/SiO$_2$ interface \cite{Tosi2017} (Fig. \ref{fig:Raman}a). The confining potential for this dot can be defined by metallic gates on top of the dielectric \cite{Veldhorst2014}, by the donor Coulomb potential \cite{Calderon2006} or by a mixture of both \cite{Tosi2017,Harvey-Collard2017}. The donor and the interface dot represent two possible binding sites for the electron charge, with a tunnel coupling $V_t$ between the two sites. In this case, the electron position can be treated as a two-level charge qubit, represented by the orbital Hamiltonian:

\begin{equation} \label{eq:H_orb}
\mathcal{H}_{\rm orb}=\frac{V_t}{2}\sigma_x - \frac{eE_{\rm dc}d}{2}\sigma_z,
\end{equation}
where $\ket{d}$ is the charge state describing the electron at the donor, $\ket{i}$ the state at the interface dot (Fig. \ref{fig:Raman}a), and $\sigma_z = \ket{d}\bra{d} - \ket{i}\bra{i}$. $d$ is the vertical distance between the centers of mass of the donor and dot wavefunctions, $e$ is the electron charge. A static vertical electric field $E_{\rm dc}$ shifts the relative energy of the donor and dot basis states. Here we define $E_{\rm dc}=0$ as the value of vertical electric field where the electron is equally shared between the donor and the interface dot.

At this biasing point, the hyperfine coupling $A$ is strongly sensitive to electric fields, with $\partial A=\partial E \approx 10^4$~Hz$\cdot$m/V for $V_{\rm t}$ of the order of 10 GHz \cite{Tosi2017}. Here, directly applying Eq.~\ref{eq:pns} would yield a nuclear spin electric dipole $p_E^{\rm ns} \sim 0.2$~Debye. However, in this regime the quantum mechanical nature of the charge degree of freedom needs to be taken fully into account. As we will now show, this results in an even stronger nuclear spin electric dipole.

At $E_{\rm dc}=0$, the charge qubit energy splitting is simply the tunnel coupling $V_t$, and the charge eigenstates are symmetric and antisymmetric combinations of $|d\rangle$ and $|i\rangle$. For arbitrary $E_{\rm dc}$, the energy splitting between the ground $\lvert g\rangle$ and excited $\lvert e\rangle$ charge eigenstates is:

\begin{equation} \label{eq:e_o}
\epsilon_{\rm o}=\sqrt{\left(V_t\right)^2+\left(eE_{\rm dc}d\right)^2}
\end{equation}

Accordingly, the hyperfine coupling energy becomes explicitly dependent on the operator $\sigma_z$ that describes the state of the charge:

\begin{equation} \label{eq:H_A}
\mathcal{H}_{A}^{\rm orb}=\frac{A}{2}\left(1 - \sigma_z\right)~{\bf S\cdot I}.
\end{equation}

This results in a coupling $g_{\rm so}$ between the flip-flop and charge qubits that takes the simple form:

\begin{equation} \label{eq:g_so}
g_{\rm so}=\frac{A}{4}\frac{V_t}{\epsilon_{\rm o}}.
\end{equation}

When $E_{\rm dc} \approx 0$, i.e. then the electron is at the ``tipping point'' between donor and interface dot, the coupling between charge and flip-flop qubits becomes as strong as $A/4$. 

Here we have assumed that the hyperfine coupling between electron spin and $^{31}$P nucleus is purely isotropic \cite{Feher1959}, i.e. dominated by the Fermi contact hyperfine term. This assumption may no longer exactly hold when the donor electron wave function is distorted from its spherical symmetry in the presence of the strong vertical electric field, whereby a small dipolar component can be created (a related case, where the electron is shared between two proximal P donors, has been recently studied \cite{Hile2018}). However, it is known that the Fermi contact component of the hyperfine coupling for donor is silicon is always the dominant term, even for $^{29}$Si nuclei which are placed off-center with respect to the symmetry point of the wavefunction \cite{Ivey1975}. Therefore, we expect the isotropic approximation to capture the main physics of the problem.

Fig. \ref{fig:Raman}c shows the level diagram for a nuclear spin Raman transition taking the electron charge levels into account. We assume that the charge and flip-flop qubit energy splittings are detuned by an amount $\delta_{\rm so} = \epsilon_{\rm o} - \epsilon_{\rm ff}$, and that an AC electric field $E_{\rm ac}\cos{(2\pi\nu_Et)}$ at frequency $\nu_E$ is applied in addition to $E_{\rm dc}$. This introduces a coupling between the charge qubit states $\ket{g}$ and $\ket{e}$ described by the rate:

\begin{equation} \label{eq:g_E}
g_{E}=\frac{e E_{\rm ac} d}{4}\frac{V_t}{\epsilon_{\rm o}}.
\end{equation}
We wish to operate in a detuned regime where $\delta_{\rm so} = \epsilon_{\rm o} - \epsilon_{\rm ff} \gg g_{\rm so}$, in which case $\ket{e}$ is minimally excited whereas $A$ remain strongly sensitive to electric fields ($E_{\rm dc} \approx 0$). In this case, the coupling between flip-flop qubit and AC electric field becomes:
\begin{equation}
g_E^{\rm ff}\approx \frac{g_Eg_{\rm so}}{2}\left(\frac{1}{\delta_E}+\frac{1}{\delta_{\rm so}}\right),
\end{equation}
which corresponds to a strong electric-dipole flip-flop transition ($\sim80$~Debye, assuming $\delta_{\rm so}=10g_{\rm so}$). Combining the strong flip-flop drive with the magnetic drive at rate $g_B$ (Eq.~\ref{eq:g_E^ns_simple}) results in a strong electric dipole transition for the nuclear spin ($\sim8$~Debye, assuming $\delta=10g_B$). However, since the nuclear transition frequency depends linearly on $A$ (see Eq.~\ref{eq:epsilon_ns}), and $A$ is a very sensitive function of electric field near the donor ionization point, electrical noise in the device will cause fast dephasing of the nuclear precession.

\begin{figure}
	\centering
	\includegraphics[width=\columnwidth]{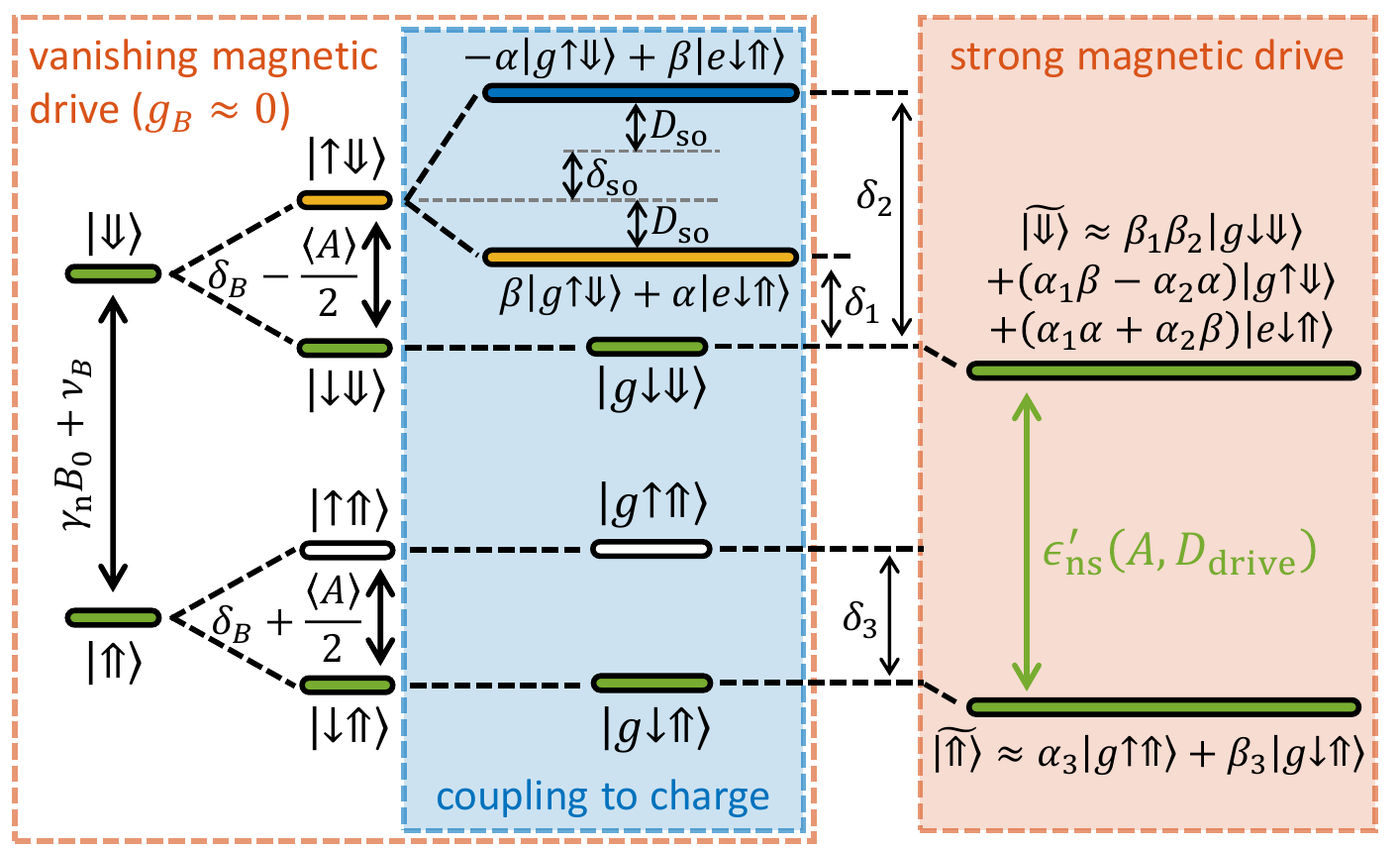}
	\caption{
 		Nuclear spin levels in the rotating frame of the magnetic drive $B_{\rm ac}$ at frequency $\nu_B$. From left to right, the system eigenstates are shown while adding the electron spin state, then the charge state, then increasing the strength of the magnetic drive. See main text for a detailed description.}
	\label{fig:levels}
\end{figure}

\begin{figure}
\centering
	\includegraphics[width=0.9\columnwidth]{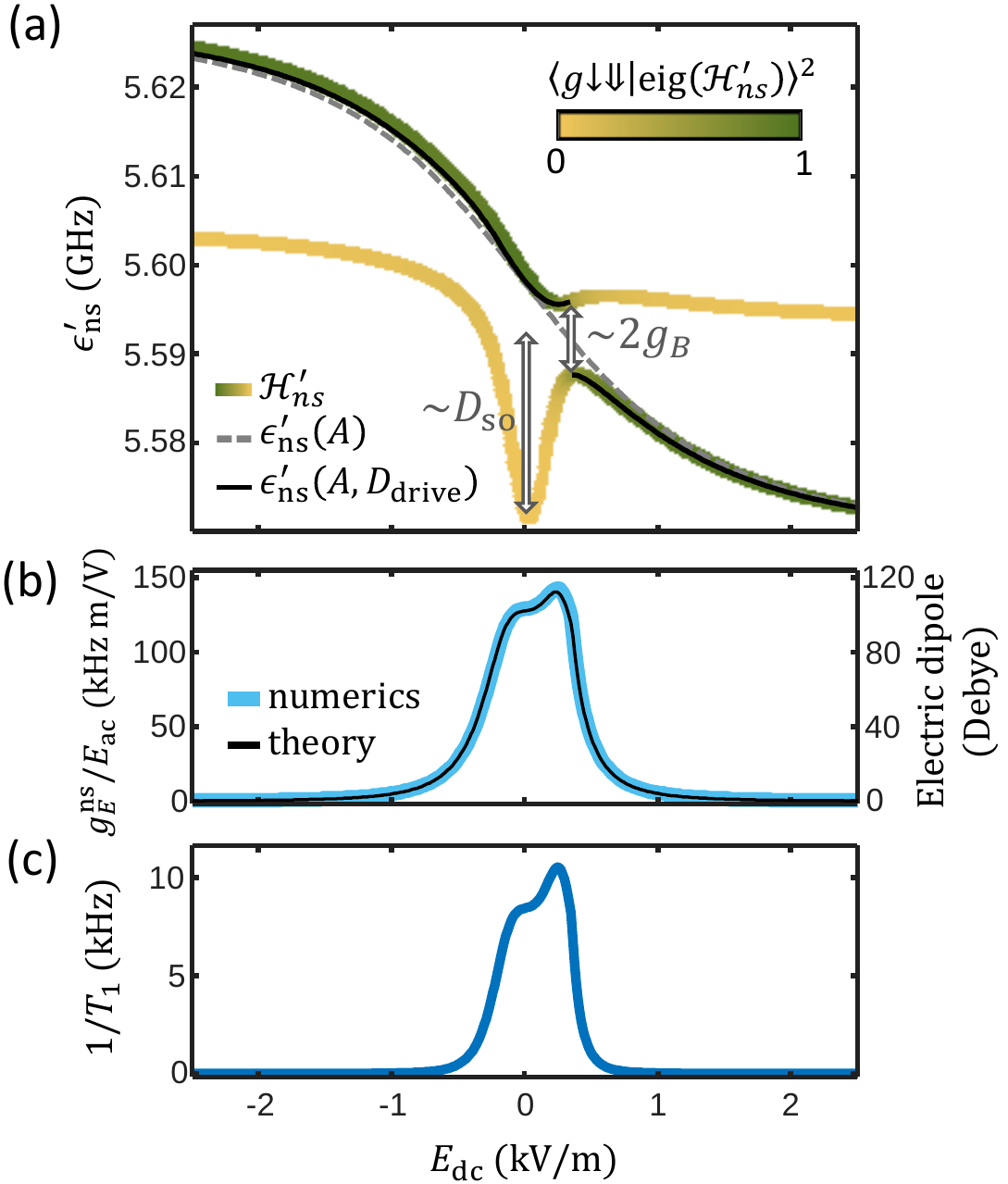}
	\caption{
(a), Nuclear spin transition frequency $\epsilon'_{\rm ns}$ in the rotating frame of the magnetic drive $B_{\rm ac}$, as a function of the static vertical electric field $E_{\rm dc}$ across the donor-dot system, for vanishing magnetic drive ($\epsilon'_{\rm ns}(A)$ -- Eq.~\ref{eq:e_ns}, grey dashed line) and strong magnetic drive ($\epsilon'_{\rm ns}(A,D_{\rm drive})$ -- Eq.~\ref{eq:e_ns_Ddrive}, black solid line). We have assumed $B_0=0.2$~T, $B_{\rm ac}=0.6$~mT, $d=15$~nm and $V_t\approx \epsilon_{\rm ff}$ and $\nu_B\approx\gamma_eB_0-A/4$ (since $\langle A \rangle = A/2$ at the ionization point). Green/yellow lines show transition frequencies calculated numerically from the Hamiltonian in Eq. \ref{eq:H_nsRot}. The color indicates the degree of admixture of the bare $\ket{g\downarrow\Downarrow}$ state into the higher $\mathcal{H}'_{ns}$ eigenstate corresponding to each transition. The nuclear spin transition (predominantly $\ket{g\downarrow\Uparrow}\leftrightarrow\ket{g\downarrow\Downarrow}$) anticrosses a flip-flop transition (predominantly $\ket{g\downarrow\Uparrow}\leftrightarrow\ket{g\uparrow\Downarrow}$) at $E_{\rm dc}=350$~V/m, with a splitting $\sim2g_B$ set by the strength of the magnetic drive. The flip-flop transition is strongly shifted by $D_{\rm so}$, due to its coupling to the charge qubit states around $E_{\rm dc}=0$. At $E_{\rm dc}=250$~V/m, the nuclear-spin excited eigenstate has $\sim75\%$ of $\ket{g\downarrow\Downarrow}$ and is robust against electrical noise ($\partial\epsilon'_{\rm ns}/\partial E_{\rm dc}=0$).
		(b), Nuclear electric dipole strength $p_E^{\rm ns}=\partial g_E^{\rm ns} / \partial E_{\rm ac}$ obtained from Eqs.~\ref{eq:g_E^ns} (theory, black line), or for numerical diagonalization of the full Hamiltonian $\mathcal{H}'$ under $E_{\rm ac}$ drive (numerics, light blue line). For the choice of parameters used in this figure,  $p_E^{\rm ns}$ peaks where $E_{\rm dc}=250$~V/m. 
		(c). Nuclear spin relaxation rate $1/T_{\rm 1,ns}$ in the presence of the magnetic drive $B_{\rm ac}$ and the effect of coupling to phonons via charge states, Eq.~\ref{eq:T1ff}.
	}
	\label{fig:clock}
\end{figure}

\section{Robust electric dipole transition of a Si:P nuclear spin} \label{sec:robust}

\subsection{Electron, nuclear and charge hybridization} 

We now show that, by adopting a different choice of device tuning, the nuclear spin can be made largely insensitive to electrical noise, while having its electric dipole transition increased even further. This is achieved by tuning the charge qubit in resonance with the flip-flop qubit ($\epsilon_{\rm o} \approx \epsilon_{\rm ff}$, i.e. $\delta_{\rm so}\approx 0$), the magnetic drive in resonance with the electron spin ($\delta_B = \gamma_eB_0 - \langle A \rangle /2 -\nu_B\approx 0$), and the electric drive in resonance with the flip-flop (and charge) qubit ($\delta_E \approx 0$). In this strongly hybridized regime, second-order perturbation theory can not be directly applied. We therefore analyze the nuclear spin Hamiltonian by first making it time-independent. The total Hamiltonian of the system is:
\begin{equation}
\mathcal{H} = \mathcal{H}_{B_0} +\mathcal{H}_{A}^{\rm orb} + \mathcal{H}_{\rm orb} + \mathcal{H}_{\rm ESR},
\end{equation}
where $\mathcal{H}_{\rm ESR}=B_{\rm ac}\cos(2\pi\nu_Bt)$ describes the magnetic drive. Next we express $\mathcal{H}$ in the rotating frame of the magnetic drive by using the transformation:
\begin{subequations}
\begin{equation}
\mathcal{H}'=U^\dagger\mathcal{H}U-i\hbar U\dot{U}^\dagger,
\end{equation}
\begin{equation}
U=e^{i2\pi\nu_Bt\left(S_z+I_z\right)}.
\end{equation}
\end{subequations}
After dropping counter-rotating terms, the transformed Hamiltonian becomes time-independent:

\begin{multline} \label{eq:H_nsRot}
\mathcal{H}'=\delta_BS_z-(\gamma_nB_0+\nu_B)I_z\\
+\frac{B_{ac}}{2}\left(\gamma_eS_x-\gamma_nI_x\right)+\mathcal{H}_{\rm orb}+\mathcal{H}_A^{\rm orb},
\end{multline}

The dominant energy scale in the above Hamiltonian is given by the term $-(\gamma_nB_0+\nu_B)I_z$, which represents the energy splitting of the nuclear spin states, but shifted to microwave frequencies by the transformation to the rotating frame of $\mathcal{H}_{\rm ESR}$. The corresponding energy levels are shown as $\ket{\Uparrow}, \ket{\Downarrow}$ at the left-most end of Fig.~\ref{fig:levels}. These levels are further split by the electron spin Hamiltonian, $\left(\delta_B+\langle A\rangle I_z\right)S_z+2g_BS_x$, where the expectation value of the hyperfine coupling $\langle A\rangle$ depends on the electron charge state, yielding the electron-nuclear spin levels shown in Fig. \ref{fig:levels}, depicted in the limit of vanishing $B_{\rm ac}$ (and therefore $g_B$). In this case, the nuclear-spin transition frequency, in the rotating frame, with the electron in the ground state, is:

\begin{equation} \label{eq:e_ns}
\epsilon'_{\rm ns}(A) = \gamma_nB_0 + \nu_B + \langle A\rangle/2.
\end{equation}

In Fig. \ref{fig:clock}a we plot (dashed line) $\epsilon'_{\rm ns}(A)$ by including the dependence of $\langle A \rangle$ on vertical electric field $E_{\rm dc}$ (from Eqs.~\ref{eq:H_orb},\ref{eq:H_A}), and is valid when the electron charge states are far detuned from the spin levels ($\delta_{\rm so}\gg g_{\rm so}$), as discussed in the previous section. The plot highlights the strong dependence of $\epsilon'_{\rm ns}$ on electric fields under such conditions. 

However, the nuclear spin dispersion changes dramatically when $\delta_{\rm so}$ approaches zero. In that case, $\mathcal{H}_A^{\rm orb}$ hybridizes the flip-flop and charge states, as shown in the blue panel within Fig. \ref{fig:levels}. The overall ground state is $\ket{g\downarrow\Uparrow}$, but the excited flip-flop state splits into two hybridized states $\beta \ket{g\uparrow\Downarrow}+ \alpha\ket{e\downarrow\Uparrow}$ and $-\alpha \ket{g\uparrow\Downarrow} + \beta\ket{e\downarrow\Uparrow}$, with:

\begin{subequations} \label{eq:alphabeta}
\begin{eqnarray}
\alpha=\frac{\theta}{\sqrt{\theta^2+1}},~~~\theta=\frac{\delta_{\rm so}-\sqrt{{\delta_{\rm so}}^2+4{g_{\rm so}}^2}}{2g_{\rm so}},
\end{eqnarray}
\begin{eqnarray}
\beta=\frac{\phi}{\sqrt{\phi^2+1}},~~~\phi=\frac{\delta_{\rm so}+\sqrt{{\delta_{\rm so}}^2+4{g_{\rm so}}^2}}{2g_{\rm so}},
\end{eqnarray}
\end{subequations}
so that $\alpha=\beta=1/\sqrt{2}$ for $\delta_{\rm so}=0$. 

As a final step, by increasing the magnetic drive amplitude $B_{\rm ac}$, the Hamiltonian term $2g_BS_x$ couples the electron spin $\ket{\uparrow}$ and $\ket{\downarrow}$ states, further hybridizing the system eigenstates. Two of those eigenstates, which we call $\widetilde{\ket{\Downarrow}}$ and $\widetilde{\ket{\Uparrow}}$ (Fig.~\ref{fig:levels}, orange box), are chiefly composed of the tensor product of the nuclear $\ket{\Downarrow}$, $\ket{\Uparrow}$ states with the ground charge state $\ket{g}$ and the ground $\ket{\downarrow}$ electron spin state. They are obtained as:

\begin{subequations} \label{eq:tildestates}
\begin{multline}
\widetilde{\ket{\Downarrow}}\approx \beta_1 \beta_2\ket{g\downarrow\Downarrow} +\\
+ (\alpha_1 \beta - \alpha_2 \alpha \ket{g\uparrow\Downarrow} + (\alpha_1 \alpha + \alpha_2 \beta)\ket{e\downarrow\Uparrow},
\end{multline}
\begin{equation}
\widetilde{\ket{\Uparrow}}\approx \alpha_3 \ket{g\uparrow\Uparrow} + \beta_3\ket{g\downarrow\Uparrow},
\end{equation}
\end{subequations}
with coefficients $\alpha_i, \beta_i$ ($i=1,2,3$) given by:

\begin{subequations} \label{eq:alphaibetai}
\begin{equation} \label{eq:alpha1}
\alpha_1=\frac{\theta_1}{\sqrt{{\theta_1}^2+1}},~~~
\theta_1=\frac{\delta_1-\sqrt{{\delta_1}^2+(2{\beta g_B})^2}}{2\beta g_B}
\end{equation}
\begin{equation} \label{eq:beta1}
\beta_1=\frac{\phi_1}{\sqrt{{\phi_1}^2+1}},~~~
\phi_1=\frac{\delta_1+\sqrt{{\delta_1}^2+(2{\beta g_B})^2}}{2\beta g_B}
\end{equation}
\begin{equation} \label{eq:alpha2}
\alpha_2=\frac{\theta_2}{\sqrt{{\theta_2}^2+1}},~~~
\theta_2=\frac{\delta_2-\sqrt{{\delta_2}^2+(2\alpha g_B)^2}}{2\alpha g_B}
\end{equation}
\begin{equation} \label{eq:beta2}
\beta_2=\frac{\phi_2}{\sqrt{{\phi_2}^2+1}},~~~
\phi_2=\frac{\delta_2+\sqrt{{\delta_2}^2+(2{\alpha g_B})^2}}{2\alpha g_B}
\end{equation}
\begin{equation} \label{eq:alpha3}
\alpha_3=\frac{\theta_3}{\sqrt{{\theta_3}^2+1}},~~~
\theta_3=\frac{\delta_3-\sqrt{{\delta_3}^2+4{g_B}^2}}{2g_B}
\end{equation}
\begin{equation} \label{eq:beta3}
\beta_3=\frac{\phi_3}{\sqrt{{\phi_3}^2+1}},~~~
\phi_3=\frac{\delta_3+\sqrt{{\delta_3}^2+4{g_B}^2}}{2g_B}
\end{equation}
\end{subequations}

The energy splitting between $\widetilde{\ket{\Downarrow}}$ and $\widetilde{\ket{\Uparrow}}$, $\epsilon'_{\rm ns}$, equals the bare nuclear-spin transition, $\epsilon'_{\rm ns}(A)$ (Eq.~\ref{eq:e_ns}), plus an amount that dependents on $E_{\rm dc}$:

\begin{equation} \label{eq:e_ns_Ddrive}
\epsilon'_{\rm ns}(A,D_{\rm drive})=\epsilon'_{\rm ns}(A)-D_{\rm drive}(E_{\rm dc}),
\end{equation}
where $D_{\rm drive}$ is an AC-Stark shift given by:
\begin{subequations}
\begin{equation} \label{eq:Ddrive}
D_{\rm drive}(E_{\rm dc})=\sum_{i=1,2,3}\frac{\delta_i}{2}\left(\sqrt{1+\left(\frac{2g_i}{\delta_i}\right)^2}-1\right),
\end{equation}
\begin{equation} \label{eq:g_alpha_beta}
g_1=\beta g_B,~~~~g_2=-\alpha g_B,~~~~g_3=g_B.
\end{equation}
\end{subequations}

This equation agrees with numerical simulations of the full Hamiltonian of Eq.~\ref{eq:H_nsRot} (Fig. \ref{fig:clock}a). Around the ionization point, the flip-flop transition (itself strongly affected by the hybridization with the charge state) anticrosses the nuclear spin transition (in the rotating frame), creating a region where $\partial\epsilon'_{\rm ns}/\partial E_{\rm dc}=0$, i.e. a first-order `clock transition' \cite{Bollinger1985,Wolfowicz2013} where $\epsilon'_{\rm ns}$ is insensitive to electric noise to first order. Further adjustment of the parameters allows for $\partial^2\epsilon'_{\rm ns}/\partial {E_{\rm dc}}^2=0$ (second-order clock transition), improving noise insensitivity even further.

In a key result of our proposal, the small admixture of the excited charge state, $\ket{e}$, into $\widetilde{\ket{\Downarrow}}$ creates an electric-dipole transition for the nuclear spin. Indeed, the $\widetilde{\ket{\Downarrow}} \leftrightarrow \widetilde{\ket{\Uparrow}}$ transition can be electrically-driven at a rate given by the charge admixture coefficients in Eq.~\ref{eq:alphaibetai} (see also Fig.~\ref{fig:clock}b):

\begin{equation} \label{eq:g_E^ns}
g_{E}^{\rm ns}=g_E\beta_3\left(\alpha_1\alpha+\alpha_2\beta\right),
\end{equation}

This electric dipole transition, at microwave frequencies, can reach $>100$~Debye around $E_{\rm dc}=0$ (Fig.~\ref{fig:clock}c). This means that even an extremely weak AC electric field, $E_{\rm ac}\approx 3$~V/m, can drive a nuclear spin transition at a MHz Rabi frequency. This is two orders of magnitude faster than the typical Rabi frequencies obtained with standard (NMR) magnetic drive at radiofrequency \cite{Pla2013}, and an order of magnitude faster than obtained (at very high electric drive amplitudes) in a recent experiment where electrically-driven NMR was achieved by modulating the quantization axis of the electron spin \cite{Sigillito2017}.

\subsection{Resilience against charge noise}

The issue of charge noise is of paramount importance in semiconductor spin qubits. It is known, experimentally and theoretically, that charge fluctuators yield a $1/\nu$ frequency dependence of the noise spectral density \cite{Paladino2014}. These models capture the averaged collective effect of many charge fluctuators on the qubit operation. In this case, charge noise results in a slow drift of the qubit electrostatic environment. Indeed, since individual qubit operations take less than a microsecond, the qubit environment is usually static within a single operations, but fluctuates in between operations. Experimentally, average quasi-static charge detuning noises around 1-9~$\mu$eV are typically found in a range of semiconductor nanodevices, including Si/SiGe \cite{Kim2015,Thorgrimsson2017,Freeman2016}, GaAs/AlGaAs \cite{Dial2013} and Si/SiO$_2$ \cite{Harvey-Collard2017,Freeman2016,Chan2018}. In particular, MOS structures were found recently \cite{Freeman2016} to have a charge noise spectrum similar to Si/SiGe devices, around $(0.5~{\rm \mu eV})^2/\nu$. Integrating over a quasi-static bandwidth relevant to qubit manipulation time-scales, e.g. between 1~Hz and 1~MHz, yields a 1.7~${\rm \mu eV}$ r.m.s. noise amplitude, expressed as a local fluctuation of electrochemical potential. In our system, given the distance between donor and interface $d \approx 15$~nm, this potential fluctuation would correspond to an r.m.s. noise on the amplitude of the vertical electric field of order 100~V/m. Inserting this into our model of the qubit energy yields a predicted nuclear spin dephasing rate of order $1-10$~kHz. Note that, similarly to dressed states \cite{London2013,Laucht2016,Laucht2017}, the addition of the strong magnetic drive has the effect of extending the coherence of our qubit. However, here the suppressed noise is of electrical nature (despite the drive being magnetic), given the particular hybridization with charge states.

We thus derived the striking result that the nuclear spin has a strong electric dipole despite being robust against electrical noise. This is because, while the qubit precession frequency is insensitive to noise, its effective transverse matrix element is strongly dependent on electric fields. Importantly, the electric dipole is induced on the nuclear spin  only around the flip-flop transition frequency, which is at several GHz. Since the charge and gate noise in nanoscale devices mainly has a $1/\nu$ spectrum, the power spectral density of the noise at the frequency that would affect the nuclear qubit is expected to be very weak. Moreover, at the same bias point where the `clock transition' ($\partial\epsilon'_{\rm ns}/\partial E_{\rm dc}=0$) for the nuclear energy takes place, the nuclear electric dipole itself is also first-order insensitive to electrical noise, since $\partial g_{E}^{\rm ns}/\partial E_{\rm dc}=0$ (Fig. \ref{fig:clock}b). A realistic $1.5~\mu$eV charge detuning noise \cite{Freeman2016} would make $g_{E}^{\rm ns}$ fluctuate by only $\sim2\%$. In other words, in this system both the free precession frequency and the Rabi frequency can be made first-order insensitive to charge noise.

As a final note, although we assumed $\delta_{\rm so} \rightarrow 0$, the electric and magnetic driving fields are still off-resonance with the eigenstates of the full Hamiltonian $\mathcal{H}_{\rm ns}$ (hybridized charge-flip-flop states), ensuring minimal excitation of the $\ket{\uparrow}$ and $\ket{e}$ states.

\subsection{Coupling to microwave cavity photons}

This strong electric dipole at microwave frequencies provides a pathway for strongly coupling $^{31}$P nuclear spins to microwave resonators \cite{Blais2004}, where a vacuum field $E_{\rm vac}$ of a few V/m can result in vacuum Rabi splittings around 1~MHz. This could be achieved \textit{e.g.} by connecting the top blue gate on Fig.~\ref{fig:Raman}a to the center pin of a superconducting coplanar waveguide resonator. Our proposal thus provides a solution to the fact that the standard (NMR) nuclear-spin transition does not naturally couple to microwave resonators. Similarly to other proposals  \cite{Pachos2002,Childress2004,Feng2008,Abanto2010}, here it is a classical drive ($B_{\rm ac}$) that enables coupling to a quantum field ($E_{\rm vac}$).

\subsection{Nuclear spin relaxation}

The engineered nuclear electric dipole also opens up a new pathway for nuclear spin relaxation: $\widetilde{\ket{\Downarrow}}$ can decay into $\widetilde{\ket{\Uparrow}}$ through a peculiar effect, where a photon from the driving field is combined with the nuclear spin energy (which is at radiofrequency) to emit a phonon at microwave frequency. The rate for this process can be roughly estimated as the admixture of the $\ket{e}$ charge excited state into the $\widetilde{\ket{\Downarrow}}$ eigenstate times the charge relaxation rate ${1}/{T_{1,\rm o}}$:

\begin{equation}\label{eq:T1ff}
\frac{1}{T_{1,\rm ns}}=\frac{|\widetilde{\bra{\Downarrow}}e\rangle|^2}{T_{1,\rm o}}\approx \frac{|\alpha_1\alpha+\alpha_2\beta|^2}{{T_{1,\rm o}}},
\end{equation}

where ${1}/{T_{1,\rm o}}=\Theta\epsilon_{\rm o}{V_t}^2$ (Ref.~\onlinecite{Boross2016}), with $\Theta\approx2.37\times10^{-24}~{\rm s}^2$ determined by the silicon crystal properties. 

As Fig.~\ref{fig:clock}c shows, ${1}/{T_{1,\rm ns}}$ peaks, around the ionization point, at a value that is still two orders of magnitude slower than e.g. the spin's coupling rate to a microwave resonator, therefore allowing the strong coupling regime to be well within reach.

\begin{figure*}
	\centering
	\includegraphics[width=0.8\textwidth]{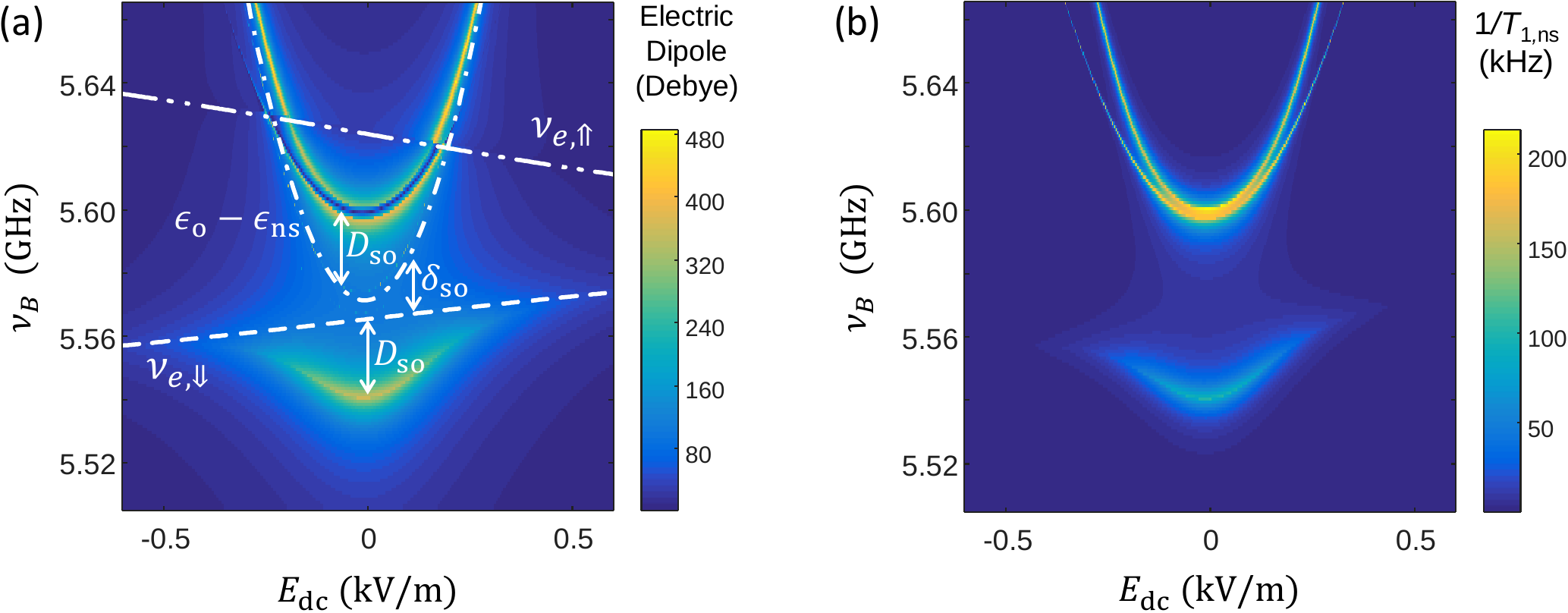}
	\caption{
		(a) Nuclear electric dipole strength $p_E^{\rm ns}$ and
		(b) nuclear spin relaxation rate $1/T_{\rm 1,ns}$, as a function of the donor-dot electric field detuning, $E_{\rm dc}$, and the magnetic drive frequency, $\nu_B$. $E_{\rm dc}=0$ is the ionization point, where the electron charge is equally shared between donor and interface dot. In (a), the dashed line shows the ESR frequency $\nu_{e,\Downarrow}$ when the nuclear spin is in the  $\ket{\Downarrow}$ state, the dot-dashed line shows the charge qubit frequency minus the nuclear spin frequency, $\epsilon_{\rm o} - \epsilon_{\rm ns}$, and the dot-dot-dashed line the electron spin resonance frequency $\nu_{e,\Uparrow}$ when the nuclear spin is in the $\ket{\Uparrow}$ state. The charge and flip-flop states are detuned by $\delta_{\rm so}$, which is close to zero at $E_{\rm dc}=0$. Charge and flip-flop states then hybridize, shifting the system eigenenergies by an AC-Stark shift $D_{\rm so}$. The plots in Figs.~\ref{fig:clock}b,c correspond to specific line cuts of the graphs shown here, for $\nu_B=\nu_{e,\Downarrow}$ at $E_{\rm dc}=0$, \textit{i.e.} $\nu_B=5.565$~GHz.
	}
	\label{fig:fig4}
\end{figure*}

\subsection{Dependence of electric dipole strength and spin relaxation rate on frequency and field detuning} \label{supp:detdep}

In Fig.~\ref{fig:clock} we have shown an operation point ($E_{\rm dc}=250$~V/m and $\nu_B=5.565$~GHz) where the proposed nuclear spin electric dipole transition is robust against noise, i.e. both its precession frequency, $\epsilon_{\rm ns}'$, and electric dipole strength, $p^E_{\rm ns}=g_E^{\rm ns}/E_{\rm ac}$, are to first order insensitive to small perturbations of the static electric field. To understand how the system behaves when slightly detuned from the optimal working point, we calculated the dependence of the nuclear spin electric dipole strength $p^E_{\rm ns}$ and relaxation rate $1/T_{1, \rm ns}$ on the magnetic drive frequency, $\nu_B$, and on the static electric field, $E_{\rm dc}$. 

The results are plotted in Fig. \ref{fig:fig4}. Both plots show two branches (bright yellow) where both dipole moment and relaxation rate are enhanced. To understand these branches, we refer to the level diagrams in Figs. \ref{fig:clock}b,c. First, note that $\nu_B$ unequivocally sets the electric dipole transition frequency $\nu_E$ (in the simplest case, $\nu_E=\nu_B+\epsilon_{\rm ns}$). The two bright branches in  Fig.~\ref{fig:fig4}a,b correspond to $\nu_E$ being in resonance with either of the two charge-flip-flop hybridized states (yellow and blue states inside the light blue rectangle in Fig. \ref{fig:levels}). If the charge and flip-flop states were uncoupled or off-resonance, then the lower branch would simply correspond to the flip-flop dipole transition, $\nu_E=\nu_{e,\Downarrow}+\epsilon_{\rm ns}$ (where $\nu_{e,\Downarrow}$ is the electron spin resonance frequency when the nuclear spin is in the `down' state), which means that the magnetic drive frequency simply coincides with the electron spin resonance $\nu_B=\nu_{e,\Downarrow}$. This would represent a simple, on-resonance Raman transition, i.e. as in the sketch in Fig.~\ref{fig:Raman}b but where $\delta=0$. Then, the upper branches in Fig.~\ref{fig:fig4} would correspond to the pure charge transition, $\nu_E=\epsilon_{\rm o}$, or equivalently $\nu_B=\epsilon_{\rm o}-\epsilon_{\rm ns}$. However, since the charge and flip-flop states are coupled, they hybridize and further split the two branches by an amount equal to $D_{\rm so}$.

Upon closer inspection, the upper branch shows an extra subtle feature. This branch correspond to excitation conditions that put the magnetic drive frequency close to the electron spin resonance frequency when the nuclear spin is in the $\ket{\Uparrow}$ state, $\nu_{e,\Uparrow}$. This, in turn, creates a pair of dressed electron spin states that further split the upper branch into two, separated by the ESR (magnetic) Rabi frequency of the $\nu_{e,\Uparrow}$ resonance. 

\section{Long-distance coupling of nuclear spin qubits} \label{sec:long}

\begin{figure}
\centering
\includegraphics[width=\columnwidth]{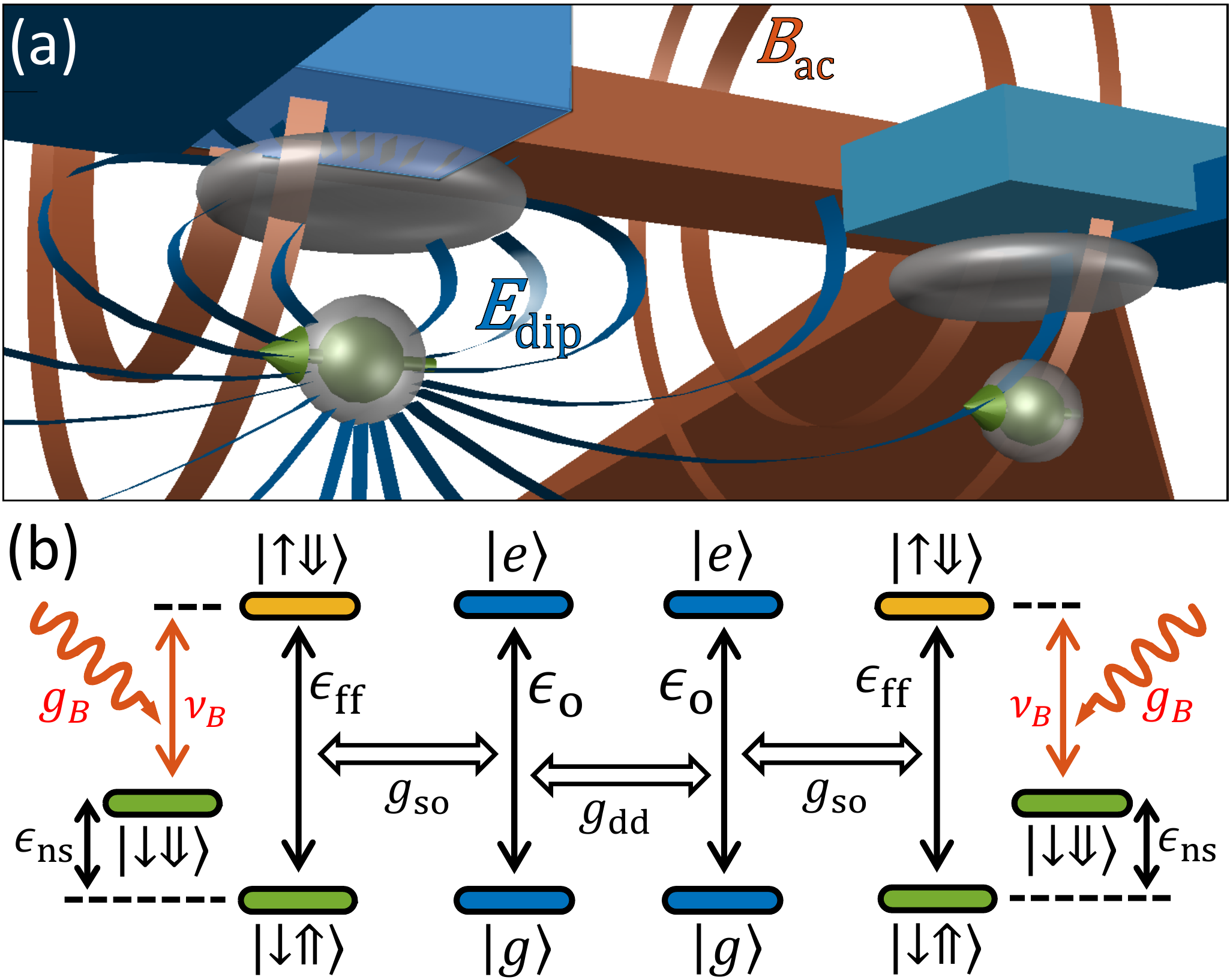}
\caption{
(a), Components and (b) level diagram for long-distance coupling of two $^{31}$P nuclear spins via electric dipole-dipole interactions. Each displaced electron produces an electric dipole field $E_{\rm dip}$ (shown only for one electron). The charge dipoles induced by displacing the electron wavefunction partly towards the interface dot interact with a strength $g_{\rm dd}$ (Eq.~\ref{eq:g_dd}), and the charge qubits interact with the flip-flop states with strength $g_{\rm so}$ (Eq.~\ref{eq:g_so}). Adding the (global) magnetic drive of strength $g_B$ and tuning the system to the fully-hybridized regime described in Sec.~\ref{sec:robust} results in a nuclear-nuclear coupling strength $g_{\rm 2q}^{\rm ns} \approx 0.55$~MHz at a $400$~nm distance (Eq.~\ref{eq:nucSWAP}).
}
\label{fig:2-qubit_nuc}
\end{figure}

We have shown in the previous section that a robust electric dipole at microwave frequencies is induced on the nuclear spin by the magnetic drive $B_{\rm ac}$, combined with the spin-charge hybridization that is obtained by displacing the electron from the donor towards an interface quantum dot. A natural and important extension of this effect is to exploit the induced electric dipole to achieve a long-distance coupling of the nuclear spins, mediated by long-range electric dipole interaction (Fig. \ref{fig:2-qubit_nuc}a) \cite{Tosi2017}. 

The interaction energy between two charge dipoles, $\textbf{p}_1=e\textbf{d}_1$ and $\textbf{p}_2=e\textbf{d}_2$, separated by $\textbf{r}$, is:

\begin{equation} \label{eq:V_dd_3D}
V_{\rm dip}=\frac{e^2}{4\pi\varepsilon_0\varepsilon_r}\frac{\textbf{d}_1\cdot\textbf{d}_2-3(\textbf{d}_1\cdot\textbf{r})(\textbf{d}_2\cdot\textbf{r})/r^2}{r^3},
\end{equation}

where $\varepsilon_0$ is the vacuum permittivity and $\varepsilon_r$ the material's dielectric constant ($\varepsilon_r=11.7$ in silicon). For the donor-dot system under study here, the electric dipole $p_{\rm o}$ depends on the operator $\sigma_z$ describing the charge state of system as: 
\begin{equation}
p_{\rm o,k} = ed_k\frac{1+\sigma_{z,k}}{2},~~~~k=1,2.
\end{equation}
The dipole-dipole interaction of Eq.~\ref{eq:V_dd_3D} thus results in a coupling term between the charge qubits (Fig. \ref{fig:2-qubit_nuc}b) equal to  \cite{Tosi2017}:
\begin{equation} \label{eq:g_dd}
g_{\rm dd}=V_{\rm dip}\frac{V_{t,1}V_{t,2}}{4\epsilon_{\rm o,1}\epsilon_{\rm o,2}}.
\end{equation}

Two distant nuclear spin qubits can then be coupled when both electrons are around their ionization point, and an AC magnetic drive $B_{\rm ac}$ is applied (Fig. \ref{fig:2-qubit_nuc}a,b) to each of them, resulting in the electric dipole $p_E^{\rm ns}$ at microwave frequencies. For the operation parameters used in Fig.~\ref{fig:clock}, $\epsilon_{\rm o}\approx\epsilon_{\rm ff}\approx\nu_B+\epsilon_{\rm ns}$ and $g_B\ll g_{\rm so}$, the two-qubit coupling rate is obtained, via second-order perturbation theory, as:

\begin{equation} \label{eq:nucSWAP}
g_{\rm 2q}^{\rm ns}=\left(\frac{g_B}{g_{\rm so}}\right)^2g_{\rm dd},
\end{equation}

which is valid if $g_B\ll (g_{\rm so})^2/g_{\rm dd}$. For two nuclear spins $r=400$~nm apart, $g_{\rm 2q}^{\rm ns}=0.55$~MHz, yielding a $\sqrt{i\mathrm{SWAP}}$ gate time of $\sim230$~ns. To put this in perspective, the Kane's proposal \cite{Kane1998} described a system of  two $^{31}$P nuclear spins placed $r=15$~nm apart, where a $\sqrt{i\mathrm{SWAP}}$ gate mediated by the electron spin exchange interaction requires $3~\mu$s -- an order magnitude slower, for over an order of magnitude tighter spacing. A recent proposal by Hill et al. describes a CNOT gate between nuclear spins mediated by the electron magnetic dipole interaction \cite{Hill2015}, wherein the 2-qubit gate time requires 300~$\mu$s for donors spaced $30$~nm apart -- three orders of magnitude slower than the electric-dipole mediated gate we have introduced here.

This method of coupling nuclear spin qubits at long distances via their induced electric dipole can be switched off completely -- $p_E^{\rm ns} \approx 0$ when the electron charge is moved back to the donor -- thus offering great flexibility in how multi-qubit operations are undertaken in a large array of qubits. The magnetic drive $B_{\rm ac}$ necessary to induce the dipole can be a global, always-on field, acting on every donor in the array. This can be optimally achieved by placing the device in a three-dimensional microwave cavity with good $B_{\rm ac}$ homogeneity \cite{Angerer2016}. Alternatively, $B_{\rm ac}$ could be delivered locally using a grid of microwave striplines \cite{Li2018}. The ``robust'' mode of operation described in Sec.~\ref{sec:robust} requires $\delta_B \approx 0$, i.e. $B_{\rm ac}$ in resonance with the electron spin transition. However, this resonance condition must be met while the donor is at the ionization point, where the hyperfine couping is approximately half the value it has while the electron is fully at the donor ($\langle A\rangle \approx A/2$), thus $\nu_B \approx \gamma_e B_0 - A/4$. Therefore, idle qubits with the electron resting at the donor will be left unaffected by the global magnetic drive, and completely decoupled from both electric and magnetic AC-fields.

\section{Conclusion} \label{sec:conclusion}

The exceptional quantum coherence of $^{31}$P nuclear spins in isotopically enriched $^{28}$Si is experimentally well established \cite{Saeedi2013,Muhonen2014}. However, it has been widely accepted that using the $^{31}$P nuclear spin as the physical qubit in a quantum computer architecture requires dealing with the very small nuclear magnetic dipole, which renders operation and multi-qubit coupling slow and cumbersome \cite{Kane1998,Ogorman2016,Hill2015}, even with inter-donor spacings $\sim 10$~nm. Indeed, most of the recent focus on $^{31}$P nuclei for quantum information has been on using them as long-lived quantum memories \cite{Morton2008,Freer2017} rather than data qubits. 

By engineering an electric dipole transition, we have shown here that the $^{31}$P qubit can also be driven at microwave frequencies, and coupled to other nuclei or to microwave cavities via electric dipole interactions, thus making it also a convenient system as data qubit. The effects of electrical noise can be strongly suppressed by operating around `clock transitions', which allow the $^{31}$P system to retain dephasing times in the $0.1 - 1$~ms range. The nuclear spin, equipped with an artificial electric dipole, can then be incorporated into large hybrid quantum architectures \cite{Xiang2013} where -- in analogy to flip-flop qubits \cite{Tosi2017} -- large arrays of nuclear qubits couple either by electric dipole-dipole coupling or via cavity microwave photons. In such architectures, the spacing between qubits can be several hundreds of nanometers, leaving ample space for classical interconnects \cite{Fischer2015,Franke2018} and readout devices, fabricated using conventional silicon nanoelectronics fabrication methods.

\begin{acknowledgments}
We thank A. P{\'a}lyi for insightful comments. This research was funded by the Australian Research Council Centre of Excellence for Quantum Computation and Communication Technology (CE110001027 and CE170100012), the US Army Research Office (W911NF-13-1-0024 and W911NF-17-1-0200) and the Commonwealth Bank of Australia.
\end{acknowledgments}


\begin{thebibliography}{53}
\bibitem [{cou()}]{coupling}%
  \BibitemOpen
  \href@noop {} {}\bibinfo {note} {We define all the coupling rates $g$ in a
  manner consistent with the Jaynes-Cummings Hamiltonian, where $g$ corresponds
  to half the Rabi frequency in the one-photon limit. Furthermore, in every
  case of resonant driving we assume a linearly polarized field, resulting in a
  Rabi frequency that equals half the dipole matrix element times the driving
  field amplitude (rotating-wave approximation). This explains the factors of 4
  appearing in all the formulas for coupling rates.}\BibitemShut {Stop}%
\end{thebibliography}

\end{document}